\documentclass[%
reprint,
superscriptaddress,
groupedaddress,
 amsmath,amssymb,
prstab,
]{revtex4-2}

\usepackage{dblfloatfix}
\usepackage{graphicx}
\usepackage{dcolumn}
\usepackage{bm}
\usepackage{siunitx}
\usepackage{placeins}
\usepackage{xcolor}
\usepackage{upgreek}

\usepackage{hyperref}
\hypersetup{
    colorlinks=true,
    linkcolor=blue,
    filecolor=blue,      
    urlcolor=blue,
    citecolor=blue,
}

\begin{document}

\preprint{APS/123-QED}

\title{Coherent High-Harmonic Generation with Laser-Plasma Beams}

\author{S.~A.~Antipov}\email{sergey.antipov@desy.de}
\author{I.~Agapov}
\author{R.~Brinkmann}
\author{{\'A}.~Ferran~Pousa}
\author{A.~Martinez~de~la~Ossa}\email{alberto.martinez.de.la.ossa@desy.de}
\author{E.~A.~Schneidmiller}
\author{M.~Th\'evenet}

\affiliation{Deutsches Elektronen-Synchrotron DESY, Notkestr. 85, 22607 Hamburg, Germany}

\date{\today}

\begin{abstract}

Active energy compression scheme is presently being investigated for future laser-plasma accelerators. This method enables generating laser-plasma accelerator electron beams with a small, $\sim 10^{-5}$, relative slice energy spread. When modulated by a laser pulse, such beams can produce coherent radiation at very high, $\sim 100$-th harmonics of the modulation laser wavelength, which are hard to access by conventional techniques. The scheme has a potential of providing additional capabilities for future plasma-based 
facilities 
by generating stable, tunable, narrow-band radiation.


\end{abstract}

\pacs{29.27.--a, 29.27.Bd}   
\keywords{Plasma acceleration, synchrotron radiation, harmonic generation}
\maketitle

\section{Introduction}

Nowadays, synchrotron radiation is playing an important role in many fields of science and engineering: from medicine and biology to chemistry, physics, and material science. Synchrotron radiation sources are offering bright, tunable photon beams with wavelengths ranging from sub-{\AA}ngstr{\"o}m to hundreds of nanometers. At the same time, there is a significant interest in employing compact and energy-efficient sources like laser-plasma accelerators (LPAs), which are making rapid progress towards practical user applications. Both plasma-based injectors~\cite{PIPIV_CDR,Panofski:2022kdk,Antipov,Shi:2022auk} and Free Electron Laser (FEL) facilities~\cite{Wang2021,Labat2023,Assmann2020} are presently being designed and built.

While there are multiple ways of producing coherent radiation at a desired wavelength from an electron beam, LPA beams come with peculiar properties that make many conventional schemes technically challenging. For example, in a Self-Amplified Spontaneous Emission (SASE)~\cite{SALDIN2000185} FEL scheme a combination of low emittance, high peak current, and low energy spread is required, otherwise the FEL gain length may become too long for practical applications.
Recently, the first breakthrough experiments have been done to demonstrate SASE at 27~nm~with an LPA~\cite{Wang2021}. Yet, jitters and energy spread in the LPA beam might affect the quality of the photon radiation in terms of its bandwidth and wavelength stability. The energy spread and jitter can be drastically reduced by employing an active energy compression scheme~\cite{Antipov, PhysRevLett.129.094801} that trades off the reduction in energy variation against a decompression in time. Consequently, achieving sufficiently low energy spreads may lead to insufficient peak beam currents, resulting in unacceptably long gain lengths. 
In principle, a low-gain FEL~\cite{Kim:2018hzb} could avoid this limitation by taking advantage of an optical cavity to build up radiation over multiple passages of the electron beam, thus decreasing the gain length. But due to the relatively low, $< 100$~Hz LPA repetition rates this scheme seems to be unfeasible.

\newpage



Alternatively, synchrotron radiation can be enhanced by modulating the density of the beam~\cite{Csonka:1975xj}, for example using a seed laser as demonstrated in~\cite{Labat2023}. 
However, the emission wavelength is typically limited by that of the seed laser to the visible spectrum.
In order to significantly reduce the emission wavelength, one may consider high-harmonic generation~\cite{ARTAMONOV1980247, PhysRevA.44.5178, gallardo1990optical, PhysRevLett.53.2405,10.1063/1.34627}. In this scheme, the electron beam is modulated by a laser pulse in the first undulator, passes through a dispersive chicane that converts energy modulation into longitudinal density modulation, and radiates in the last undulator. The key factor limiting the access to higher harmonics in this scheme is the uncorrelated energy spread within the bunch $\sigma_E$. Efficient generation of harmonics requires an energy modulation amplitude $A = \Delta E / \sigma_E \geq n$, making generation of high harmonics $n \gg 1$ particularly challenging. 


Active energy compression allows the LPA beam reaching extremely low levels of energy spread and jitter, down to $\sim 10^{-5}$, employing conventional radiofrequency (RF) accelerating cavities for active energy compression~\citep{Antipov}. This scheme is envisioned for the PETRA~IV plasma injector~\cite{PIPIV_CDR} to enable clean and efficient injection in the storage ring. Once filled, the storage ring needs to be topped up only once every few minutes to restore a small fraction ($\sim 1$~\%) of the total charge lost due to Touschek and residual gas scattering~\footnote{Contemporary and planned 4th generation light sources, such as PETRA~IV feature large $\gtrsim 10$~h beam lifetimes.}. 
This opens a window of opportunity for additional applications of an LPA injector that could take advantage of  high quality electron beams, when they are not required for the storage ring.
Thanks to their small energy spread after energy compression the LPA beams require a relatively small energy modulation to access high $\sim 100$-th harmonics of the seed wavelength, while a high power laser for the seeding is naturally present in the LPA setup. Thus, the LPA electron beams can produce stable, narrow-band radiation at high harmonics of the driving laser pulse, achieving wavelengths as small as 10~nm for an 800~nm seed.



\begin{figure*}[!t]
    \includegraphics[width=\textwidth]{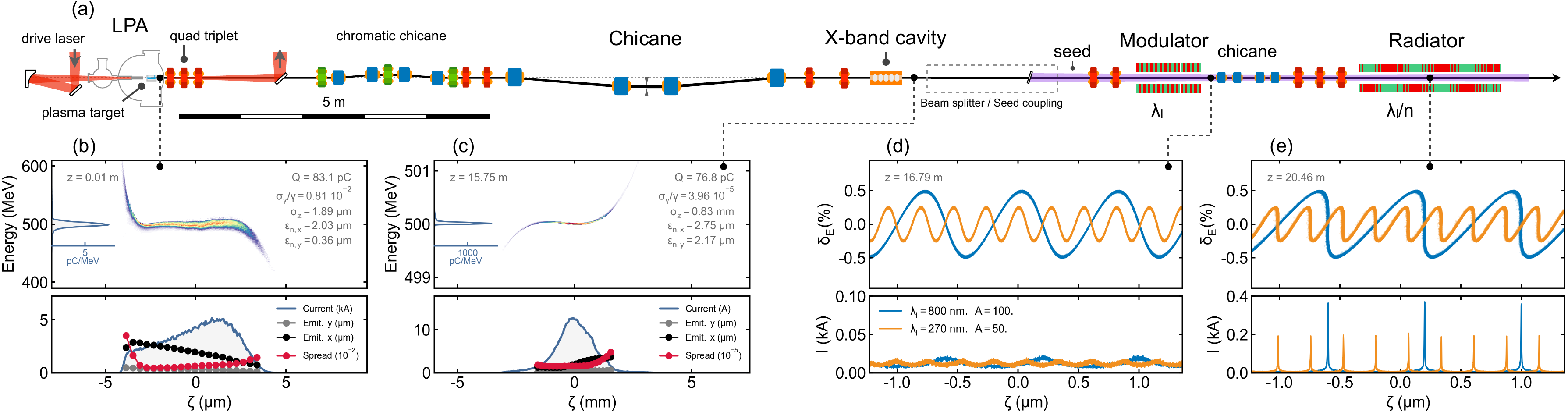}
    \caption{\label{fig:schematic}Schematic of the proposed setup (a). An LPA generates an electron bunch with an rms energy spread of 1\% (b); the bunch then passes through an energy compression beamline, which lengthens it and reduces the energy spread to $< 10^{-4}$ (c). The resulting beam is then modulated by a laser pulse (d) and forms microbunches in the final undulator where it radiates on a harmonic of the modulating wavelength (e). Numerical simulation by \texttt{FBIPC}~\cite{bib:fbpic} (LPA) and \texttt{Ocelot}~\cite{bib:ocelot} (beamline).
    }
\end{figure*}

Figure~\ref{fig:schematic}(a) shows the proposed setup, based on a 500~MeV LPA injector described in~Ref.~\citep{Antipov}. It consists of an LPA followed by a quadrupole triplet to capture the electron beam from the plasma cell, a chicane with sextupoles for chromaticity correction, a large stretcher chicane, and finally an RF cavity for suppressing energy spread and correcting energy deviations.   

\newpage
\onecolumngrid
\twocolumngrid
The injector beamline drastically reduces the relative energy spread of the beam from $\sim 10^{-2}$ to $< 10^{-4}$~[Fig.~\ref{fig:schematic}(b,c)].
The energy-compressed beam then enters the first undulator where it co-propagates with a seed laser pulse that is generated by splitting a small fraction of the drive pulse. The laser field imprints a modulation of the beam energy with a spatial period of the laser wavelength~[Fig.~\ref{fig:schematic}(d)]. After passing through a dispersive section the energy modulation translates into microbunching with the periodicity of the modulation wavelength~[Fig.~\ref{fig:schematic}(e)]. The resulting microbunches have a peak current far exceeding  that of the initial bunch. And finally, the resulting train of short microbunches radiates in the second undulator at the high $n \gg 1$ harmonic of the seed laser wavelength $\lambda_l$.


\section{Analytical estimates}

To describe the process in more detail, let us consider a Gaussian energy distribution for a longitudinally uniform beam slice prior to the modulator: $f_0(E) = \exp[-(E - E_0)^2 / 2 \sigma_E^2] / \sqrt{2 \pi} \sigma_E$, with $E_0$ and $\sigma_E$ the mean energy and its rms, respectively. Assuming that the modulator does not affect the intra-beam particles' position, $\zeta$, it only induces a sinusoidal energy modulation $E^\prime = E + \Delta E \sin(k_l \zeta)$ with $k_l = 2\pi / \lambda_l$.
The modulation amplitude created by a laser pulse in a planar undulator can be found as~\cite{Hemsing2014}
\begin{equation}\label{eq:mod_depth}
    \Delta E = e\,\sqrt{P_l Z_0 / \pi} A_{JJ} K L_u / \gamma w_0,
\end{equation}
where $e$ is the elementary charge, $\gamma = E_0 / mc^2$ is the mean Lorentz factor of the beam, $P_l$ is the modulator laser power, $w_0$ the laser beam waist size, $Z_0 \approx 377~\Omega$ the vacuum impedance, $L_u = \lambda_u N_w$ the undulator length (with $\lambda_u$ the undulator period and $N_w$ the number of periods), $K$ the undulator parameter, and $A_{JJ} = J_0(\chi) - J_1(\chi)$, $J_n$ is the $n$-th Bessel function of the first kind, and $\chi = K^2 / (4 + 2 K^2)$. 
The undulator $K$ has to be chosen such that its resonance wavelength is tuned to the laser: 
\begin{equation}\label{eq:K}
    \lambda_l = \lambda_u (K^2 / 2 + 1) / 2 \gamma^2.
\end{equation}

\begin{figure}[!h]
    \centering
    \includegraphics[width=\columnwidth]{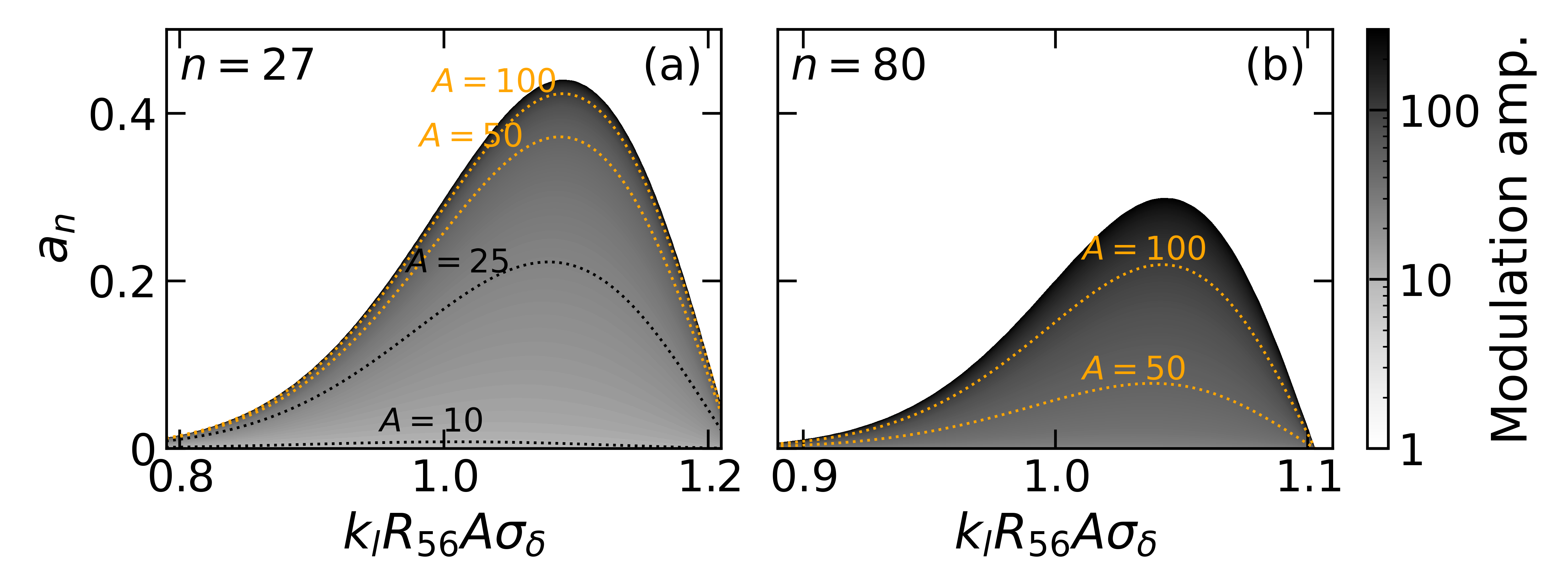}
    \caption{Bunching factors as function of the normalized dispersion parameter for different energy modulation amplitudes $A$ for the 27th (a) and the 80th (b) harmonics.}
    \label{fig:bun_factors}
\end{figure}

\begin{figure}
    \centering
    \includegraphics[width=\columnwidth]{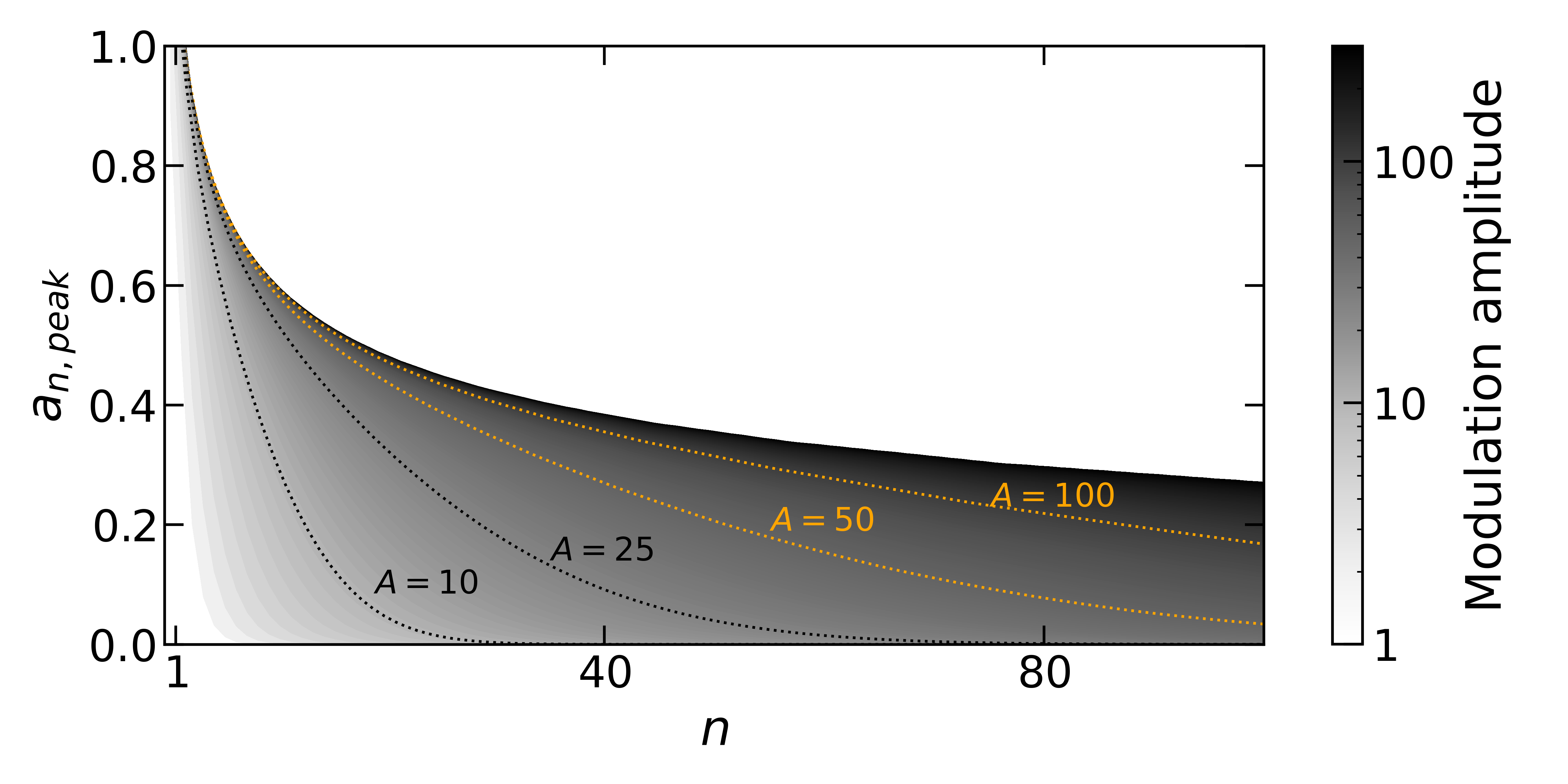}
    \caption{Peak bunching factor as a function of the harmonic number $n$ for different modulation amplitudes $A$. 
    \label{fig:max_bun_factors}}
\end{figure}

The following dispersive section is characterized (to first order) by its momentum compaction factor $R_{56}$ that shifts the particles' position $\zeta^\prime = \zeta + R_{56} (E^\prime/E_0 - 1)$. After this, the longitudinal phase space distribution of the beam becomes $f(E',\zeta') = f_0[E(E',\zeta')]$. 
The resulting modulation of the beam current is obtained as $I(\zeta') = I_0 \int_{-\infty}^{+\infty} f(E',\zeta') dE'$, with $I_0$ the initial current of the beam slice. In the limit of a uniform longitudinal density profile the modulation is periodic and 
can be expressed as a Fourier series~\citep{SALDIN2005499}
\begin{equation}
    I(\zeta') = I_0\left[ 1 + \sum_{n = 1}^\infty a_n \cos(n k_l \zeta') \right].
\end{equation}
The coefficients $a_n$ are the \emph{bunching factors} of the $n$-th harmonics of the modulation frequency, which for the phase space transformation described above can be analytically calculated as~\citep{PhysRevA.44.5178,SALDIN2005499} 
\begin{equation}\label{eq:bunching_factor}
    a_n = 2 J_n \left( n k_l R_{56} A \sigma_\delta \right) \exp \left( -n^2 k_l^2 R_{56}^2 \sigma_\delta^2 /2 \right),
\end{equation}
where $\sigma_\delta = \sigma_E / E_0$~\footnote{Note that this quantity is twice the bunching factor as it is often defined, e.g. $b_n$ in Ref.~\cite{Hemsing2014}.}. Figure~\ref{fig:bun_factors} shows them for $n=27$ and $n=80$ cases. For $n \gg 1$ the maximum is achieved when the normalized dispersion parameter
\begin{equation}\label{eq:optimum}
    k_l R_{56} A \sigma_\delta \approx 1.
\end{equation}
Thus, generation of high harmonics requires precise tuning of $R_{56}$ and strong energy modulation. When both conditions are satisfied one may achieve considerable bunching factors at high harmonics $n \gg 10$ (Fig.~\ref{fig:max_bun_factors}).

The power radiated at the $n$-th harmonic of the modulation wavelength through the last undulator is proportional to $a_n^2$ (see e.g. Ref.~\cite{Hemsing2014}). 
In order to provide an analytical estimation, we may consider the limit of a thin electron beam~\citep{SALDIN2005499}, such that the diffraction parameter $\sigma_r^2 n k_l / L_u \ll 1$. In this regime, the radiation power is independent of the transverse beam size $\sigma_r$:
\begin{equation}\label{eq:rad_power}
    W_0 = \pi^2 a_n^2 \chi A_{JJ}^2 N_w W_b I/ \gamma I_A, 
\end{equation}
where $W_b = \gamma m_e c^2 I/e$ is the total power of the electron beam and $I_A \approx 17$~kA is the Alfven current.

\section{Additional considerations}

A few additional effects might affect the idealized picture presented above. First, as the bunch passes the undulator, the undulator generates an additional energy-position correlation:
$R_{56,u} = 2 \lambda N_w$, where $\lambda$ stands for the photon wavelength, $\lambda_l$ in the modulator and $\lambda_l / n$ in the radiator.
The additional $R_{56}$ created by the modulator affects the optimum condition in Eq.~(\ref{eq:optimum}), reducing the $R_{56}$ that needs to be created by the chicane. In the radiator it affects the microbunching by displacing the beam from the optimum condition set by Eq.~(\ref{eq:optimum}) and thus limiting the distance over which the high bunching factor remains. We can estimate this distance by applying the condition $R_{56,u} / R_{56} \lesssim \Delta_a$, where $\Delta_a \approx 1.8/n^{2/3}$ is the relative full-width half-maximum of the bunching factor peak. For the 80-th harmonic it is about 0.1, while for the 27-th harmonic it is about 0.2 (Fig.~\ref{fig:bun_factors}).

Also, a longitudinal space charge can play a significant role in limiting the length where microbunching persists. Inside an undulator the longitudinal space charge is described by a reduced Lorentz factor $\gamma_r = \gamma / \sqrt{1 + K^2 /2}$~\cite{Geloni:2007ti}
and the debunching length is
    $L_{cr} = \gamma_r \sigma_r \sqrt{\gamma I_A / I_0}$~\cite{bib:Khan}. 
This limits the length of the radiator undulator to $L_{u} \leq L_{cr} \approx 10$~m.

The bunching factor can also be affected by the coherent synchrotron radiation (CSR). CSR effects in the beam delivery line before the energy compression have been studied previously~\cite{Antipov} and found to have no significant effect on the beam quality thanks to a rapid decompression of the beam. Let us estimate the magnitude of the CSR kick after the energy modulation using a simple steady-state 1D model~\cite{mayes_exact_2009}: $W_\mathrm{CSR} \approx N_b m_e c^2 r_e \kappa^{2/3} \sigma_{z,\mathrm{mb}}^{-4/3}$, where $\kappa = 0.042~$m$^{-1}$ is the bending curvature and $\sigma_{z,\mathrm{mb}}$ is the rms length of a microbunch. In the beginning of the chicane $\sigma_{z,\mathrm{mb}} \sim \lambda_l$ and $I_\mathrm{mb} \sim 5$~A, resulting in negligible $W_\mathrm{CSR} \sim 5$~keV/m. Approaching the end of the chicane $\sigma_{z,\mathrm{mb}}$ reaches $\sim \lambda_l / n$ and $I_\mathrm{mb} \sim 400$~A, the CSR becomes significant: $W_\mathrm{CSR} \approx 1.7~$MeV/m. But since the dipole is only 15~cm long, this results in only $5\times10^{-4}$ relative energy variation -- a small figure compared to the laser modulation amplitude.

Another detrimental effect would arise from a high beam divergence $\sigma_r^\prime$ when the beam propagates after being modulated in energy, which makes the particles slip by $L \sigma_r^{\prime 2} /2$, smearing out the microbunches over a propagation distance $L$. 
Although this effect can be easily mitigated by decreasing the beam divergence, it would limit the smallest beam size that can be achieved at and after the modulator.

Finally, the energy jitter of the electron beam will also affect the optimum condition given by Eq.~(\ref{eq:optimum}). For the achievable $10^{-4}$ levels of energy stability, the jitter does not play a significant role as it is much  smaller than the relative width of the bunching factor peak of $0.1-0.2$~(Fig.~\ref{fig:bun_factors}). The high energy stability of the electron beam also results in high spectral stability of the generated radiation.

\section{Numerical examples}

As an example, we consider the 500~MeV energy beam produced by the injector from Ref.~\cite{Antipov}, shown in Fig.~\ref{fig:schematic}~(a). 
First, the beam is obtained through Bayesian optimization of the LPA stage using the particle-in-cell code \texttt{FBPIC}~\cite{bib:fbpic} and the optimization software \texttt{Optimas}~\cite{FerranPousa2023,optimas}~[Fig.~\ref{fig:schematic}(b)]. 
Then, the beam is tracked through the energy compression and harmonic generation parts of the beamline in the \texttt{Ocelot}~\cite{bib:ocelot} code, which uses up to second-order transfer maps to simulate the beamline elements. CSR effects through the dispersive sections of the beamline (the chicanes) are computed by \texttt{Ocelot} using a projected 1D model~\cite{SALDIN1998158}.
Figure~\ref{fig:schematic}(c) shows the beam after the energy compression and Tab.~\ref{tab:beam_par} summarizes its most relevant parameters at this point. Figures~\ref{fig:schematic}(d)~and~\ref{fig:schematic}(e) show the result of the tracking through the energy modulator and the subsequent dispersive section for a central slice of the beam. 
Finally, the coherent radiation emitted by the beam in the radiator is computed using the code \texttt{Synchrad}~\cite{synchrad}.
Throughout this start-to-end simulation chain, the same macroparticles representing the simulated beam are transported from code to code transforming to the corresponding coordinate system.
\begin{table}[!h]
    \centering
    \caption{Beam parameters after energy compression.}
    \begin{tabular}{lrr}
        \hline\hline
         Parameter & Symbol & Value\\
         \hline
         Peak current & $I_0$ & 12~A\\
         Beam energy & $E_0$ & 500~MeV \\
         Bunch length, rms & $\sigma_z$ & 0.8~mm \\
         Rel. energy spread, rms & $\sigma_\delta$ & $5\times 10^{-5}$ \\
         Norm. emittance, rms & $\varepsilon_x,~\varepsilon_y$ & $2.8,~2.2~\mathrm{\upmu m}$ \\
         \hline \hline
    \end{tabular}
    \label{tab:beam_par}
\end{table}

For coherent high-harmonic generation, we consider two scenarios. In case I, the beam is modulated by a laser pulse with the wavelength of the LPA driver: $\lambda_l = 800$~nm, obtained by picking up a fraction of the drive laser. In case II, the modulation is performed by a laser pulse with three times shorter wavelength: $\lambda_l = 270$~nm, which can be obtained by frequency-conversion of a fraction of the driver laser~\cite{Labat2023}. In both cases, we target a final radiation wavelength of 10~nm, which corresponds to the 80-th harmonic of 800~nm and the 27-th of 270~nm. According to Fig.~\ref{fig:bun_factors}, in order to achieve a significant bunching factor at those harmonics, we select an energy modulation of $A = 100$ and $A = 50$, which corresponds to an absolute energy modulation amplitude of 2.50~MeV and 1.25~MeV, for cases I and II, respectively. This modulation requires a laser peak power of 2.3~(1.8)~GW and 48~(36)~mJ energy for case I~(II), according to Eq.~(\ref{eq:mod_depth}). This is a small fraction of the 2.45~J pulse used to produce and accelerate the electron beam in the LPA.
Table~\ref{tab:und_params} summarizes the undulator parameters. Note that the magnetic field of the undulator was decreased in case II to account for the smaller $\lambda_l$.
\begin{table}[!h]
    \centering
    \caption{Undulator parameters for cases I and (II).}
    \begin{tabular}{lcrr}
        \hline\hline
         Parameter & Symbol & Modulator & Radiator\\
         \hline
         Period & $\lambda_u$ & 74~mm & 14~mm\\
         Number of periods & $N_w$ & 15 & 160\\
         Peak field & $B$ & 0.91~(0.50)~T& 0.66~T\\
         Strength & $K$ & 6.3 (3.8) & 0.9\\
         Wavelength & $\lambda$ & 800~(270)~nm & 10~nm \\
         \hline \hline
    \end{tabular}
    \label{tab:und_params}
\end{table}

\begin{figure}[t]
    \centering
    \includegraphics[width=\columnwidth]{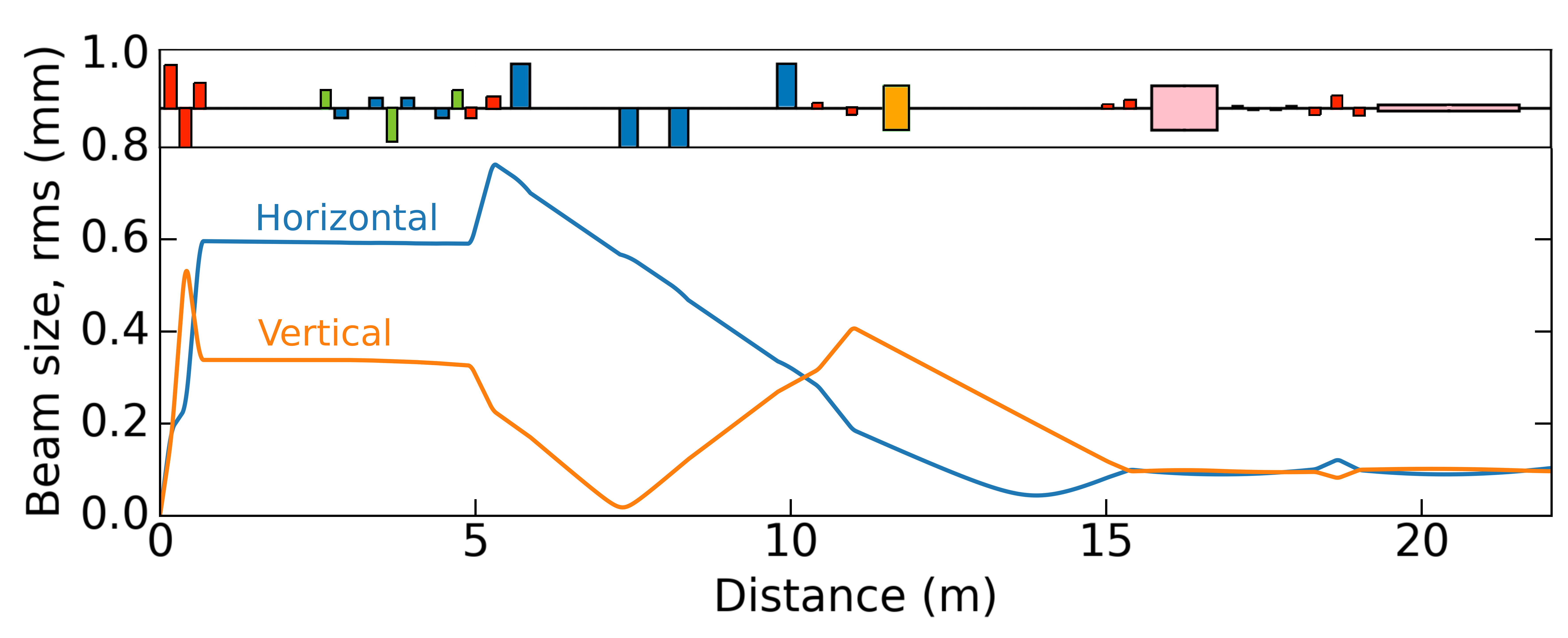}
    \caption{Betatron beam envelopes along the trajectory, computed using the \texttt{Ocelot} code~\cite{bib:ocelot}. Dipoles are shown in blue, quadrupoles in red, sextupoles in green, RF in orange, and undulators in pink.}
    \label{fig:beam_size}
\end{figure}

\texttt{Ocelot} is also used to simulate the energy modulation process in the modulator accounting for the finite size and diffraction of the laser pulse. In order to mitigate these effects a sufficiently large value for the laser waist, $w_0 = 1$~mm is selected.
Figure~\ref{fig:beam_size} shows the betatron beam sizes throughout the whole beamline. To minimize the beam divergence, these are kept approximately constant at $\lesssim 100~\upmu$m through the modulator plus chicane in both horizontal and vertical planes. 


Due to the high number of particles required to properly resolve the coherent radiation, we limit the simulation to the central slice of the beam where we track every single electron. 
Thus, a total of $4 \times 10^{6}$ particles are used to generate a flat-top beam slice of length 16~$\upmu$m and current 12~A at the entrance of the modulator. Figure~\ref{fig:schematic}(d) presents the longitudinal phase space of a central beam slice right after the modulator. 
The following magnetic chicane is tuned to maximize the bunching factor of the beam at the middle of the radiator [Fig.~\ref{fig:schematic}(e)]. For case I~(II), this results in a chicane with $R_{56} = 17~(13)~\upmu$m, just a 70\% of the value calculated from Eq.~(\ref{eq:optimum}). The additional compression is given passively by the combined effect of the modulator and radiator, and the free drifts in between. 
In case I~(II), the central slice spans a total of 20~(60) microbunches. The Courant-Snyder parameters of the beam slice differ slightly from those of the whole beam~(Table~\ref{tab:beam_par})~\footnote{In particular, the normalized emittances are 2.1 and 0.4~$\mathrm{\upmu m}$ in the horizontal and transverse planes, respectively, slightly smaller than the projected ones~[see Fig.~\ref{fig:schematic}(c)]. This results in smaller transverse sizes (rms) of the slice: $\sigma_x = 73~\upmu$m and $\sigma_y = 44~\upmu$m at the center of the radiator}. 

\begin{figure}[t]
    \centering
    \includegraphics[width=\columnwidth]{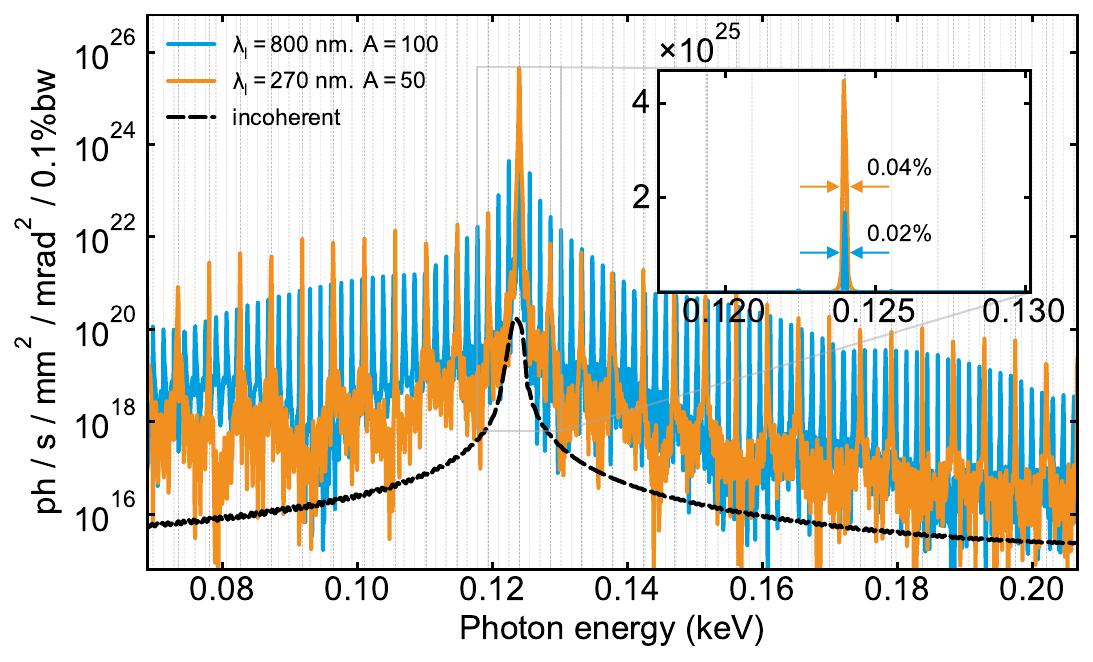}
    \caption{Peak brilliance of the radiation calculated using the code \texttt{Synchrad}~\cite{synchrad}. The blue curve shows the spectrum of the simulation case I, $\lambda_l = 800$~nm; the orange one shows the case II, $\lambda_l = 270$~nm. Vertical dashed lines depict the harmonics of the seed laser.}
    \label{fig:spectra}
\end{figure}

The radiation field emitted by the beam in the final undulator is calculated using the code \texttt{Synchrad}~\cite{synchrad}. This code computes the coherent sum of the radiation field produced individually by the particles composing the beam, using the Fourier transformed Lienard-Wiechert potentials~\cite{jackson2021classical}. To compare with the case of a non-modulated beam, we also use \texttt{Synchrad} to compute the incoherent spectrum as the sum of the radiation intensities of individual electrons.
The particle trajectories through the radiator are calculated analytically with a resolution of 64 points per undulator period, assuming a perfectly sinusoidal magnetic field and neglecting collective effects and radiation energy losses. The particle trajectories are fed into Synchrad to calculate the radiation field.
Figure~\ref{fig:spectra} shows the peak brilliance of the radiation for the two considered cases. In logarithmic scale, we can see how the coherent radiation is about five orders of magnitude greater than the incoherent signal, and characterized by the presence of narrow peaks at the harmonics of the seed. As expected, the maximum peak brilliance is reached where the fundamental radiation wavelength of the undulator is resonant with one of the harmonics, which for case I~(II) is 1.7~(4.5)~$\times 10^{25}~\mathrm{photons/s/mm^{2}/mrad^{2}/0.1\% bw}$. Figure~\ref{fig:spectra} also shows an inset in linear scale where we observe that the spectrum is strongly dominated by the resonant peak, featuring a relative rms bandwidth of $1.8\times10^{-4}$ and $3.9\times10^{-4}$ for cases I and II, respectively. The average radiated power for case I~(II) yields 1.5~(6.3)~kW. To estimate the total radiated energy, we simply integrate the result of the central slice through the longitudinal profile of the whole beam. Assuming a Gaussian temporal profile with $2.7$~ps rms~(Table~\ref{tab:beam_par}), we obtain 0.01~(0.04)~$\upmu$J for case I~(II).

\begin{figure}[t]
    \centering
    \includegraphics[width=\columnwidth]{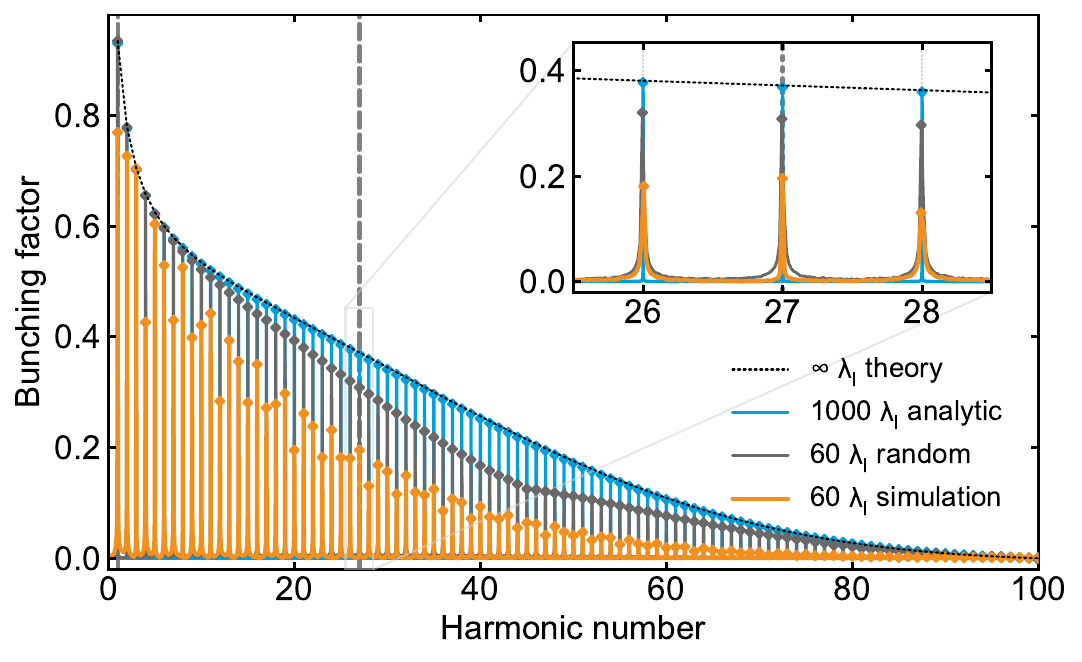}
    \caption{Bunching factor as a function of the harmonic number for a case with $A = 50$ and $k_l R_{56} A \sigma_\delta = 1.09$. The dotted line shows the theoretical value for an infinite slice according to Eq.~(\ref{eq:bunching_factor}). The blue line shows the bunching factor for an analytically modulated finite slice of length 1000~$\lambda_l$. The gray line values are obtained from a randomly generated slice of length 60~$\lambda_l$. In orange, the values calculated for the simulation case II.\label{fig:bunching_factor_comp}}
\end{figure}

We can compare the simulated radiation power with that calculated using Eq.~(\ref{eq:rad_power}). The bunching factors at the corresponding harmonics,  calculated from the Fourier transform of the current profiles of the slices, yield $a_{80} = 0.08$ and $a_{27} = 0.18$ for cases I and II, respectively. This translates into 5.1~kW and 25.6~kW for case I and II, respectively, not far from the simulated values. Note that our simulated bunching factors are somewhat smaller than the analytic limit (Fig.~\ref{fig:max_bun_factors}) due to the finite length of the simulated beam slice. Figure~\ref{fig:bunching_factor_comp} shows the calculated bunching factors for the simulated slice in case II and compares them with the analytical predictions.

\section{Conclusion}

With the help of the energy compression technique modern LPAs are approaching a point where they will be capable of delivering electron beams with low energy spread and jitter. We are proposing a scheme that takes advantage of this low energy spread to generate very high, $\sim 100$-th harmonics of the driver laser. Our estimates have shown that using a small, $\sim 1\%$ fraction of the driver laser pulse to modulate the electron beam energy one can form a train of few thousands microbunches within the full width at half maximum of the initial electron bunch. This is possible thanks to the low, $10^{-5}$-level energy spread of the electron beam. The resulting short microbunches then allow accessing high harmonics of the modulator wavelength in the radiator undulator, producing $\sim 1$-ps-long 
bursts of radiation. 
The radiation wavelength can be easily tuned by selecting the desired harmonic of the seed $\lambda_l$, adjusting the radiator $K$ parameter or the beam energy. The distance between adjacent harmonics is about $\lambda_l/n^2 \approx 0.1$~nm for the 80-th harmonic of 800~nm. Similar radiation power can be generated at each wavelength, because the bunching factor depends weakly on $n$ for high $n$'s (Fig.~\ref{fig:max_bun_factors}).
Since the width of the resulting radiation is determined by the spectrum of the beam current, the long microbunch trains would generate extremely narrow spectral lines, as seen in our numerical simulation. Ultimately, one can aim at achieving the width as small as $\sim 10^{-5}$. In practice, the spectral width and phase noise of the seed laser pulse and the nonlinear energy chirp, remaining in the electron beam after the energy compression, will broaden the spectrum. Together these effects might worsen the relative width to $\sim 10^{-4}$ level. The narrow bandwidth might find applications in spectroscopy where a high spectral resolving power is required. The proposed scheme has a potential for scalability to higher beam energies, which can help mitigating detrimental effects of space charge and beam divergence. Since the radiation wavelength is defined by the harmonics of the seed wavelength, a higher beam energy may not always result in a higher photon energy.
Still, it may enable a wider range of undulator parameters such as a longer period and higher strength, allowing for additional tunability of the resulting radiation.

\section*{Acknowledgments}
The authors would like to thank Paul Winkler, Andreas Maier, Christoph Lechner, Sergey Tomin, Igor Andriyash and other colleagues for useful discussions on laser systems, numerical simulation of coherent radiation and its applications.

\bibliography{biblio} 




\end{document}